\documentclass[aip,reprint]{revtex4-1}

\usepackage{graphicx}
\usepackage{dcolumn}
\usepackage{amsmath,amssymb}
\usepackage{bm}

\usepackage{chemformula} 
\usepackage[T1]{fontenc} 
\usepackage[colorlinks=true,allcolors=blue]{hyperref}

\newcommand*{\unit}[1]{\textrm{ #1}}
\pdfpageattr {/Group << /S /Transparency /I true /CS /DeviceRGB>>}

\begin{document}

\title[]{Observation and manipulation of a phase separated state in a charge density wave material}

\author{Sean M. Walker}
\email{s4walker@uwaterloo.ca}
\affiliation{Institute for Quantum Computing, University of Waterloo, Waterloo, Ontario, N2L 3G1, Canada}
\affiliation{Department of Chemistry, University of Waterloo, Waterloo, Ontario, N2L 3G1, Canada}
\author{Tarun Patel}
\affiliation{Institute for Quantum Computing, University of Waterloo, Waterloo, Ontario, N2L 3G1, Canada}
\affiliation{Department of Physics, University of Waterloo, Waterloo, Ontario, N2L 3G1, Canada}
\author{Junichi Okamoto}
\affiliation{Institute of Physics, University of Freiburg, D-79104 Freiburg, Germany}
\affiliation{EUCOR Centre for Quantum Science and Quantum Computing, University of Freiburg, D-79104 Freiburg, Germany}
\author{Deler Langenberg}
\author{E. Annelise Bergeron}
\affiliation{Institute for Quantum Computing, University of Waterloo, Waterloo, Ontario, N2L 3G1, Canada}
\affiliation{Department of Physics, University of Waterloo, Waterloo, Ontario, N2L 3G1, Canada}
\author{Jingjing Gao}
\author{Xuan Luo}
\author{Wenjian Lu}
\author{Yuping Sun}
\affiliation{Key Laboratory of Materials Physics, Institute of Solid State Physics, Chinese Academy of Sciences, Hefei 230031, People's Republic of China}
\affiliation{High Magnetic Field Laboratory, Chinese Academy of Sciences, Hefei 230031, People's Republic of China}
\affiliation{Collaborative Innovation Centre of Advanced Microstructures, Nanjing University, Nanjing 210093, People's Republic of China}
\author{Adam W. Tsen}
\affiliation{Institute for Quantum Computing, University of Waterloo, Waterloo, Ontario, N2L 3G1, Canada}
\affiliation{Department of Chemistry, University of Waterloo, Waterloo, Ontario, N2L 3G1, Canada}
\author{Jonathan Baugh}
\email{baugh@uwaterloo.ca}
\affiliation{Institute for Quantum Computing, University of Waterloo, Waterloo, Ontario, N2L 3G1, Canada}
\affiliation{Department of Chemistry, University of Waterloo, Waterloo, Ontario, N2L 3G1, Canada}

\begin{abstract}
The 1T polytype of \ch{TaS2} has been studied extensively as a strongly correlated system. As 1T-\ch{TaS2} is thinned towards the 2D limit, its phase diagram shows significant deviations from that of the bulk material. Optoelectronic maps of ultrathin 1T-\ch{TaS2} have indicated the presence of non-equilibrium charge density wave phases within the hysteresis region of the nearly commensurate (NC) to commensurate (C) transition. We perform scanning tunneling microscopy on exfoliated ultrathin flakes of 1T-\ch{TaS2} within the NC-C hysteresis window, finding evidence that the observed non-equilibrium phases consist of intertwined, irregularly shaped NC-like and C-like domains. After applying lateral electrical signals to the sample we image changes in the geometric arrangement of the different regions. We use a phase separation model to explore the relationship between electronic inhomogeneity present in ultrathin 1T-\ch{TaS2} and its bulk resistivity. These results demonstrate the role of phase competition morphologies in determining the properties of 2D materials.
\end{abstract}

\maketitle


Strongly correlated electronic systems exhibit rich phase diagrams often characterized by competing phases with different kinds of order.\cite{dagotto2005complexity} Charge, spin, lattice, and orbital degrees of freedom can all contribute to the existence of a variety of states exhibiting a range of bulk properties, including high-temperature superconductivity and electronic inhomogeneity.\cite{battisti2017universality,sun2021evidence,fradkin2015colloquium} Furthermore, in these complex systems, novel, emergent behaviors are observed that depend on both the properties of the phases involved and the nature of the competition between them. Depending on the balance of competition between the different states, small external stimuli have the potential to result in large changes in bulk observables,\cite{uehara1999percolative} a behavior that could be harnessed for sensing or memory devices.

The transition metal dichalcogenide (TMD) 1T-\ch{TaS2} is a van der Waals material that displays many of the features associated with complex systems. 1T-\ch{TaS2} exhibits strong electron correlations, is a Mott insulator at low temperatures, and has a rich phase diagram comprising a variety of charge density wave (CDW) phases.\cite{wilson1975charge} The material undergoes a series of successive first-order phase transitions upon cooling: starting as a normal metal, it forms an incommensurate CDW (IC) at $\sim 545 \unit{K}$, a nearly commensurate CDW (NC) at $\sim 355 \unit{K}$, and lastly a commensurate CDW (C) at $\sim 185 \unit{K}$. Each of these phases has a characteristic electronic, structural and orbital order,\cite{ritschel2015orbital} and a jump in resistivity is observed with each transition as the material becomes more insulating as it is cooled.

Moving beyond the equilibrium phases of 1T-\ch{TaS2}, the phase diagram is expanded through the application of certain external stimuli,\cite{sipos2008mott,ma2016metallic,cho2016nanoscale,stojchevska2014ultrafast,vaskivskyi2016fast} by modifying select material parameters,\cite{yu2015gate} or by thinning the material to the ultrathin limit ($ < 20 \textrm{ nm}$ thick). Previous results in ultrathin flakes of 1T-\ch{TaS2} have indicated that within the hysteresis region of the NC-C transition, the resistivity of the material can be bidirectionally switched, driven reversibly through intermediate states whose resistivities lie between the extrema values associated with the NC and C states.\cite{patel2020photocurrent,hollander2015electrically,yoshida2015memristive,ma2019observation} While this property suggests the use of this material as a memory element in future technologies, the nanoscopic nature of these non-equilibrium intermediate phases remains an open question. Efforts to characterize the CDW phases in 1T-\ch{TaS2} have typically focussed on bulk samples, within regions of the phase diagram where the CDW states are well-defined. Only recently have attempts been made to specifically study the spatial characteristics of the phases present in flakes within the NC-C hysteresis region, but these experiments have been limited to characterizing the phases at the micron length scale.\cite{patel2020photocurrent,ludwiczak2020impeded,frenzel2018infrared}

In this work, the nanoscopic nature of the CDW states in the hysteresis region of the NC-C phase transition of ultrathin 1T-\ch{TaS2} is elucidated. We image exfoliated flakes using scanning tunneling microscopy (STM) and observe the existence of an inhomogeneous electronic state. Additionally, we image changes in this state after the application of lateral electrical signals and after repeated rastering of the STM tip. The ability to tune the resistivity of the material within the NC-C phase transition region is revisited, and many of the observed features are re-examined within the context of the initial state being a phase separated state. Key characteristics of the intermediate states accessed within the NC-C phase transition are captured by a phase separation model based on time-dependent Ginzburg-Landau theory.

1T-\ch{TaS2} is known to oxidize in ambient conditions\cite{tsen2015structure} and thus previous STM experiments on bulk 1T-\ch{TaS2} have typically relied upon \emph{in situ} cleaving of crystals to expose a clean surface.\cite{wu1989hexagonal,thomson1994scanning,burk1991charge,wu1990direct} In this work, the exfoliated ultrathin samples measured were fabricated using a polymer transfer technique under an inert atmosphere, and were only exposed to air prior to being loaded into the STM ($\sim 5 \unit{min}$). While residues remained from processing the device, clean regions of interest were located and imaged by first mapping out a navigational path that avoided existing surface contamination. The STM was equipped with four electrical contacts, allowing for the resistance of the sample to be measured and for lateral electrical signals to be applied to the device.

\begin{figure*}
\centering
\includegraphics[width=0.8\linewidth]{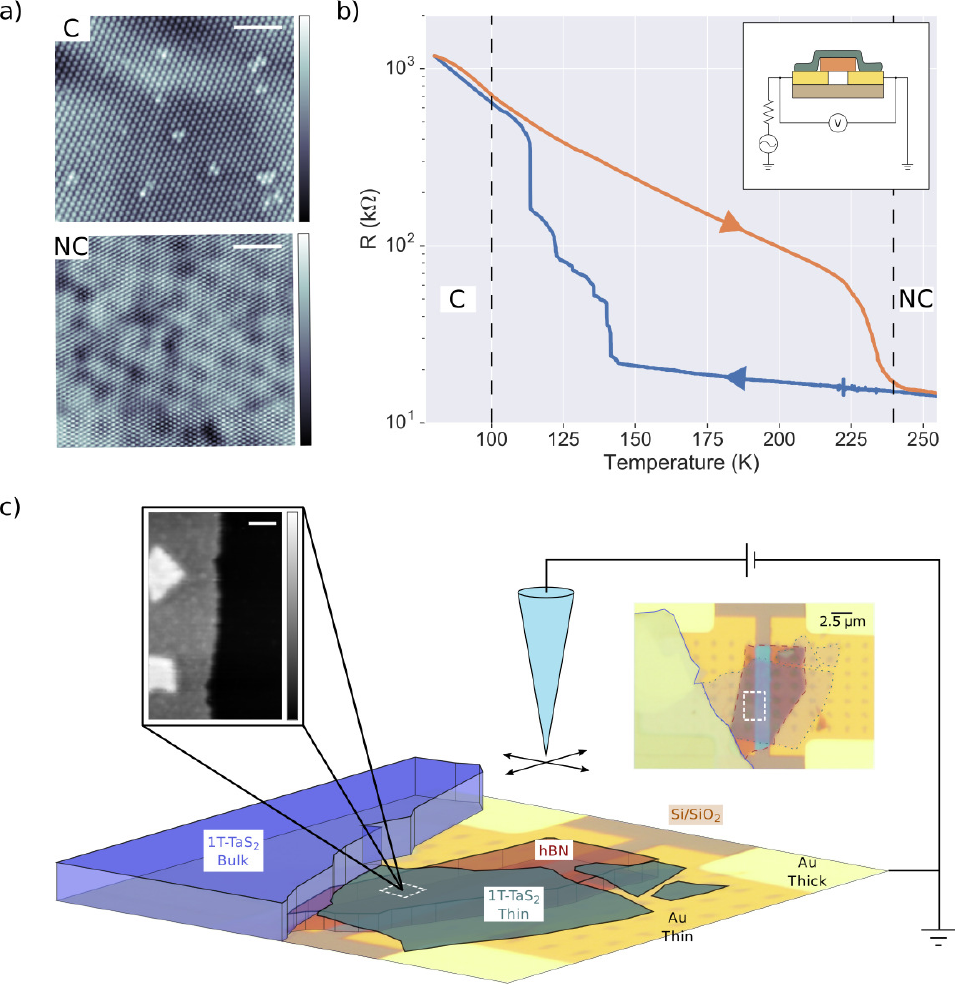}
\caption[Bulk CDW phases in 1T-TaS2 and device schematic.]{Imaging ultrathin 1T-\ch{TaS2}. (a) The C (top - scale bar $= 7 \unit{nm}$, $V_t = 0.57 \textrm{ V}$, $I_{set} = 0.19 \textrm{ nA}$) and NC (bottom - scale bar $= 11 \unit{nm}$, $V_t = 5 \textrm{ mV}$, $I_{set} = 1.5 \textrm{ nA}$) CDW phases in 1T-\ch{TaS2} measured in a bulk crystal. (b) Temperature-dependent resistance of a typical ultrathin flake of 1T-\ch{TaS2}. The orange trace is collected upon warming the sample from $77 \unit{K}$ while the blue trace is collected upon cooling. Step wise transitions are observed as the material is cooled from the NC phase to the C phase. (Inset) Cross section of the device indicating the layer stacking and the measurement geometry. The colors of the different layers match those in (c). (c) Schematic of a completed sample. The different van der Waals materials are overlayed on an optical image of the device. In this sample a bulk flake of 1T-\ch{TaS2} was transferred along with the thin flake, though only the thin flake spans the gap in the gold electrodes. The dashed white box indicates the location of a wide STM scan taken on a transferred ultrathin 1T-\ch{TaS2} flake demonstrating how the navigational markers and the gap between the gold electrodes are visible underneath the transferred flake. Scale bar $= 400 \textrm{ nm}$, $V_t = 0.75 \textrm{ V}$, $I_{set} = 0.2 \textrm{ nA}$.}
\label{fig:schematic}
\end{figure*}

Figure~\ref{fig:schematic}c illustrates the design of the samples.
Gold contacts separated by a $2 \unit{$\mu$m}$ wide gap are lithographically defined on a \ch{SiO2}/\ch{Si} substrate. Hexagonal boron nitride (\ch{hBN}) is transferred first, spanning the gap, followed by an ultrathin flake of 1T-\ch{TaS2} that makes contact with the gold pads on either side of the \ch{hBN}. The \ch{hBN} provides a substrate that is more electronically clean and atomically flat compared to the \ch{SiO2} surface. Markers written on the gold contact pads using electron beam lithography allow for navigation to regions of interest (Figure~\ref{fig:schematic}c). Initial contact in the vicinity of the flake is facilitated by patterning gold regions with different thicknesses such that the contrast is visible through a telescope attached to an optical port of the STM. With this device design it is possible to correlate changes in electronic structure with changes in the bulk electrical properties of the exfoliated flakes being measured.

\begin{figure*}
\centering
\includegraphics[width=\linewidth]{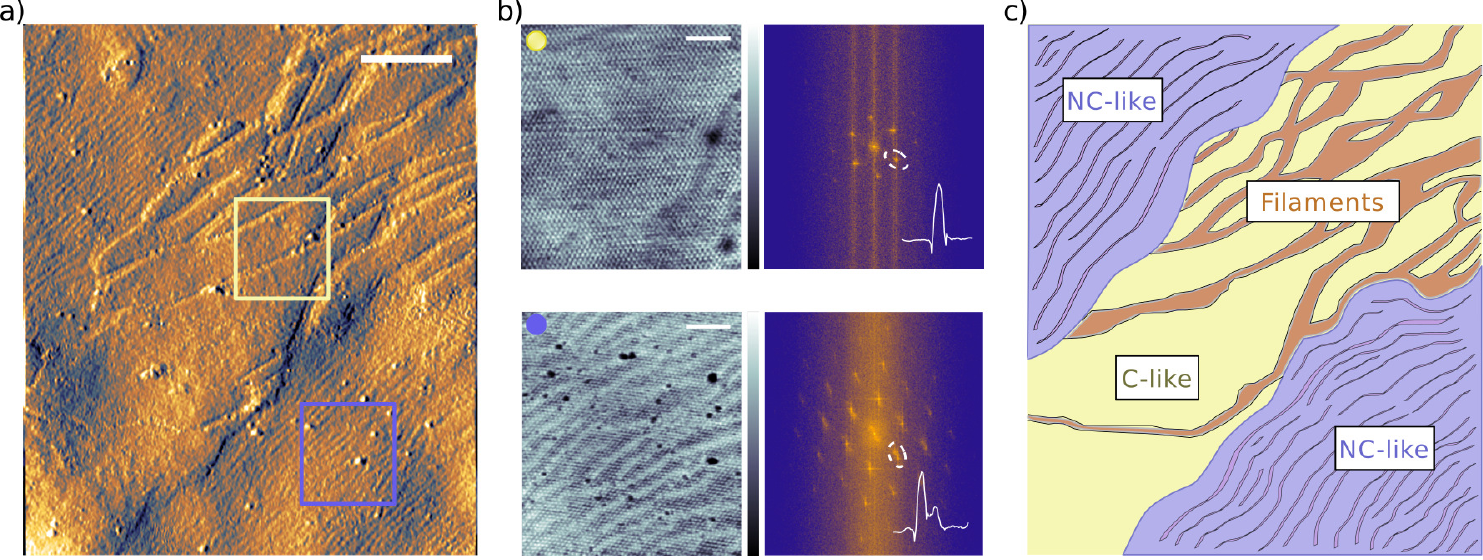}
\caption[An inhomogeneous electronic state in 1T-\ch{TaS2}.]{Inhomogeneity in ultrathin 1T-\ch{TaS2}. (a) A wide STM scan comprised of two distinct region types. In the top left and bottom right there are NC-like regions with visible striations, while a C-like region with no striations runs from the bottom left to the top right. The C region also contains filaments. The displayed image was generated by applying a Sobel transform to the flattened, raw data to extract the gradient of the scan along the vertical axis. $\textrm{Scale bar} = 70 \unit{nm}$, $V_t = 0.30 \textrm{ V}$, $I_{set} = 1.40 \textrm{ nA}$. (b) (Top) A detail of the C-like region outlined in yellow in (a), and its corresponding FFT. Only the fundamental CDW modes are visible in the FFT. A line cut through the circled peak indicates the absence of a satellite peak. $\textrm{Scale bar} = 9 \unit{nm}$, $V_t = -0.30 \textrm{ V}$, $I_{set} = 1.40 \textrm{ nA}$. (Bottom) A detail of an NC-like region similar to that outlined in purple in Figure~\ref{fig:phasecoexistence}a, and its corresponding FFT. The FFT contains the fundamental CDW modes as well as satellite peaks, as evident in a line cut through the circled peak. The presence of a satellite peak is indicative of domains. $\textrm{Scale bar} = 11 \unit{nm}$, $V_t = 0.4 \textrm{ V}$, $I_{set} = 0.6 \textrm{ nA}$. (c) A schematic delineating the different regions visible in the STM scan.}
\label{fig:phasecoexistence}
\end{figure*}

\begin{figure*}
\includegraphics[width=0.85\textwidth]{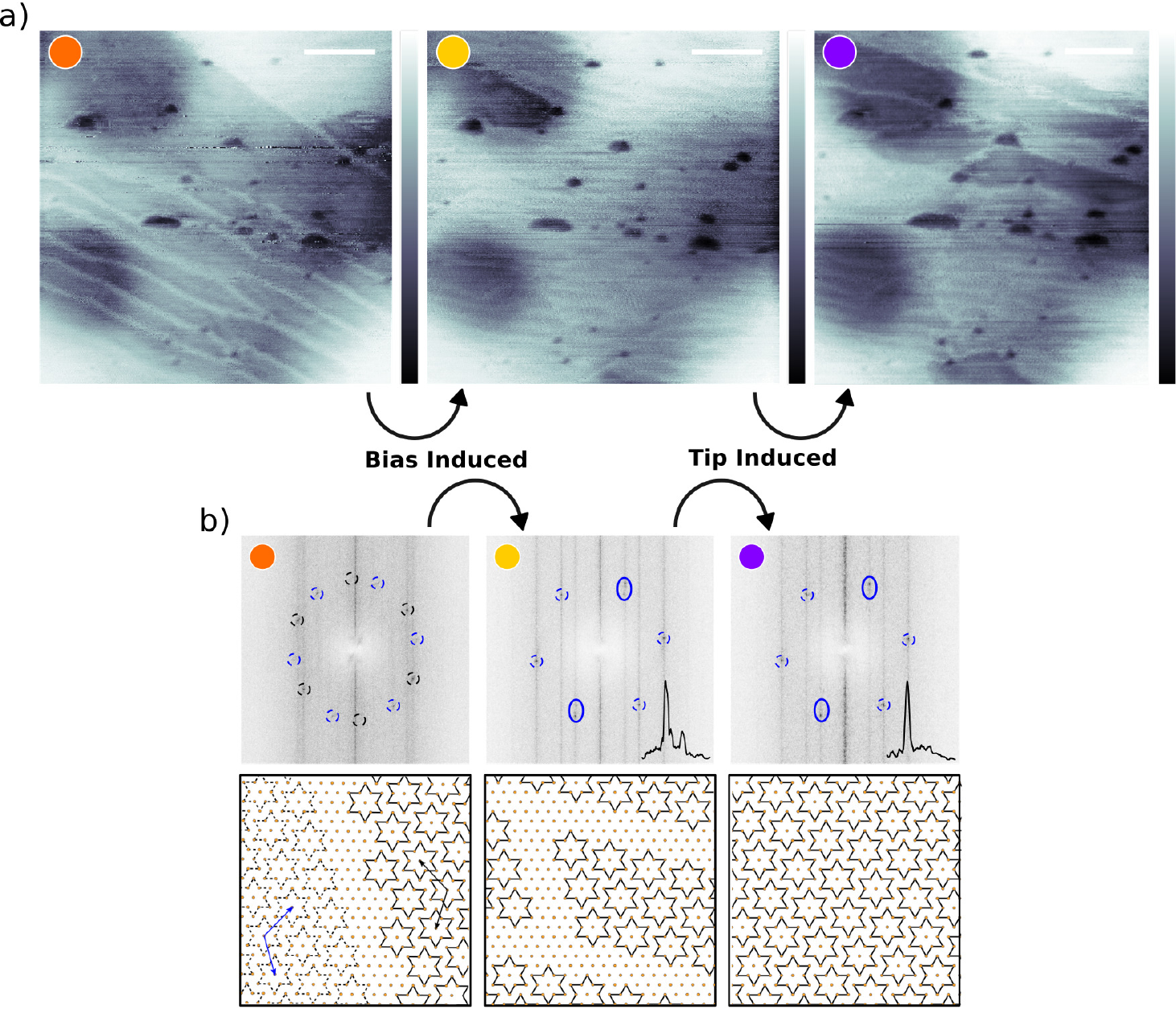}
\caption[Driving the inhomogeneous electronic state in 1T-\ch{TaS2}.]{Evolution of the inhomogeneous electronic state with the application of external stimuli. (a) STM scans illustrating the change in the electronic state with the application of an electric field and as a consequence of the rastering of the STM tip. Initially there is a mix of the C-like and NC-like phases, and the C-like region also contains filament structures. After a bias is applied across the sample, the NC region dominates. Scanning with the tip results in sections of the NC-like phase to be converted to the C-like phase. Scale bar $= 30 \unit{nm}$, $V_t = 0.30 \textrm{ V}$, $I_{set} = 1.40 \textrm{ nA}$. (b) (Top) FFTs of the STM scans in (a). The line profiles correspond to cuts through the peaks circled by a solid line, and indicate whether a satellite peak is present. (Bottom) Schematics that demonstrate how the state is evolving. The Stars of David represent the maxima of the CDW, and illustrate the clustering of the \ch{Ta} lattice that accompanies the charge ordering. Initially the state consists of a mixture of two possible chiral domains (blue and black lattice vectors), and thus $12$ peaks are observed in the FFT. After electrical driving, only one orientation remains, with the state being dominated by the NC-like phase and containing domain walls. The FFT now consists of the fundamental modes of a single orientation of the CDW with a visible satellite peak. Finally, after continuous scanning with the tip, the NC-like region has been converted to the C-like phase. Consequently, in the FFT the satellite peak has now disappeared.}
\label{fig:driving}
\end{figure*}

The data presented in Figure~\ref{fig:phasecoexistence} and Figure~\ref{fig:driving} was collected from an approximately $8 \unit{nm}$ thick flake with lateral dimensions of $5 \unit{$\mu$m} \times 10 \unit{$\mu$m}$. The sample was cooled from $260 \unit{K}$ to $170 \unit{K}$, within the NC-C transition hysteresis region, and was imaged where the flake lies on the gold contact pad and also within the conductance channel where the flake lies on \ch{hBN}. A filtered, topographic image of the flake on the gold contact pad is given in Figure~\ref{fig:phasecoexistence}a.
\twocolumngrid
On both the contact pad and the \ch{hBN}, the topography exhibits spatial inhomogeneity with two distinct region types, distinguishable by the presence or absence of bright striations. These striations are reminiscent of the domain walls present in the bulk NC phase of 1T-\ch{TaS2},\cite{wu1989hexagonal,thomson1994scanning} and thus, throughout this work, we use the label "NC-like" to identify the regions where striations are observed, and "C-like" to identify the other region type. Fast Fourier transforms (FFTs) of each region type (Figure~\ref{fig:phasecoexistence}b) support this identification. The FFTs of the NC-like regions contain satellite peaks in addition to the fundamental CDW modes. The existence of satellite peaks is indicative of the presence of domains.\cite{thomson1994scanning} In the FFT of the C-like region only the fundamental, three-fold symmetric CDW modes are observed. It is important to note that the observed state is not simply an NC state with asymmetrically arranged domain walls nor is it an NC state with a domain wall period different from the equilibrium NC value as has been previously suggested,\cite{patel2020photocurrent,ma2019observation} but rather the state is specifically composed of defined, irregularly shaped NC-like and C-like domains.

The domain walls in the NC-like region in Figure~\ref{fig:phasecoexistence}c lack three-fold symmetry and are instead similar to the discommensurations observed in the bulk T phase.\cite{tanda1984x,tanda1985x} Ginzburg-Landau theory is often used for describing a triple CDW material, like 1T-\ch{TaS2}, with wave vectors $Q^{(i)}$, $i = (1,2,3)$.\cite{mcmillan1976theory,nakanishi1977nearly} All of the bulk phases of 1T-\ch{TaS2} are known to satisfy the triple-\emph{Q} condition, $Q^{(1)} + Q^{(2)} + Q^{(3)} = 0$. While the wave vectors of the C, NC and IC phases are oriented $120^{\circ}$ degrees apart, in the T phase this constraint is relaxed.\cite{nakatsugawa2020multivalley} The fundamental CDW wave vectors in the NC-like region found in Figure~\ref{fig:phasecoexistence}c have the same characterization as the T phase, in that they satisfy the triple-\emph{Q} condition while the angles between the wave vectors are not equal to $120^{\circ}$.

To perform bidirectional resistance switching, we initially start in the low resistance state at $170 \unit{K}$, and run a series of linear sweeps of laterally applied voltage, increasing the maximum voltage set with each sweep (Figure SI~\ref{fig:SIDC}a). Figure~\ref{fig:driving}a illustrates one set of STM images taken before and after a set of electrical signals were applied. In this particular case, the sweep rate was $0.05 \textrm{ V/s}$ and the maximum voltage was $4 \textrm{ V}$. After the voltage sweeps, the measured resistance had increased by only $\sim 1\%$ (Figure SI~\ref{fig:SITASTM5}b).
Similar to Figure~\ref{fig:phasecoexistence}a, distinct NC-like and C-like regions are observed prior to applying a bias across the sample. After electrically driving the flake, the NC region now dominates the image. Considering the minimal change in the bulk resistance of the material, in this instance the driving acted to rearrange the domains of the initial microstructure.

In addition to the changes caused by the application of an electric field, we also observe interconversion between the domain types as a consequence of the rastering STM tip. The rightmost scan in Figure~\ref{fig:driving}a was taken following the electrical driving and illustrates the effect of the tip on the inhomogeneous state. With repeated scanning of the tip, at a tunneling voltage of $V_t = 0.3 \textrm{ V}$, NC-like regions of the sample within the scanning area were converted to C-like regions. Unlike what is observed during the electrical driving, the overall resistance of the sample did not change from one scan to the next, indicating that the differences in the scans were due to the local perturbation of the tip rather than being a part of a global thermodynamic process.

The full evolution of the state can be clearly seen in the changes in the FFTs of consecutive STM scans (Figure~\ref{fig:driving}b). Initially, the FFT contains 12 peaks. There are two possible chiral orientations of the CDW that arise in 1T-\ch{TaS2}, often referred to in the literature as the $\alpha$ and $\beta$ orientations.\cite{zong2018ultrafast} The initial STM scan in Figure~\ref{fig:driving}a contains a mirror domain wall separating an NC-like region of the CDW in the $\alpha$ orientation from a C-like region of the CDW in the $\beta$ orientation. Consequently, both chiral modes are observed in the FFT of the image, resulting in the appearance of 12 peaks (Figure SI~\ref{fig:SIchiral}). After lateral electrical driving of the sample, the NC region dominates the image and thus the FFT now only contains 6 peaks, representing a single orientation of the CDW. Consistent with the prevalence of the NC-like region in the image, a satellite peak in the FFT is visible. As the area is subsequently scanned, the NC region is gradually converted into the C phase, and thus in the FFT of the final scan the satellite peak is no longer present.

Given the temperature at which the STM measurements were performed, and the similarities between the features observed and the known structure of the phases present in bulk 1T-\ch{TaS2}, we believe the data indicates that the ultrathin flake is in a phase separated state. Hysteresis in an observable through a first-order phase transition is typically indicative of the presence of superheating/supercooling.\cite{chaikin1995principles} Within the hysteresis region, the two phases involved in the transition are nearly degenerate and can coexist, forming a phase separated state characterized by a complex microstructure consisting of intertwined domains of the ground state of the system and the nearly degenerate metastable state. The data presented in Figure~\ref{fig:phasecoexistence} and Figure~\ref{fig:driving} provides evidence for the presence of phase separation in ultrathin 1T-\ch{TaS2}, with the electronic state imaged comprising irregularly shaped domains of C-like and NC-like regions of a size on the order of hundreds of nanometers.

Additional evidence for the presence of phase separation in ultrathin 1T-\ch{TaS2} can be found in the thickness dependent resistivity of the material over the same temperature range. In bulk flakes, measurements of the resistivity through the NC-C transition are hysteretic. As 1T-\ch{TaS2} is thinned, the hysteresis region exhibits marked changes. In the ultrathin regime stepwise transitions are observed (Figure~\ref{fig:schematic}b), and as the $2$D limit is approached, the region widens until at a thickness of $\sim 2 \unit{nm}$ ($\sim 4$ atomic layers), the jump in the resistivity disappears altogether.\cite{tsen2015structure} This result is potentially consistent with a disorder-induced phase separated state,\cite{dagotto2001colossal} and suggests that ultrathin flakes of 1T-\ch{TaS2} cooled below the NC-C transition temperature are prone to spatial inhomogeneity.

\begin{figure*}
\includegraphics[width=0.75\linewidth]{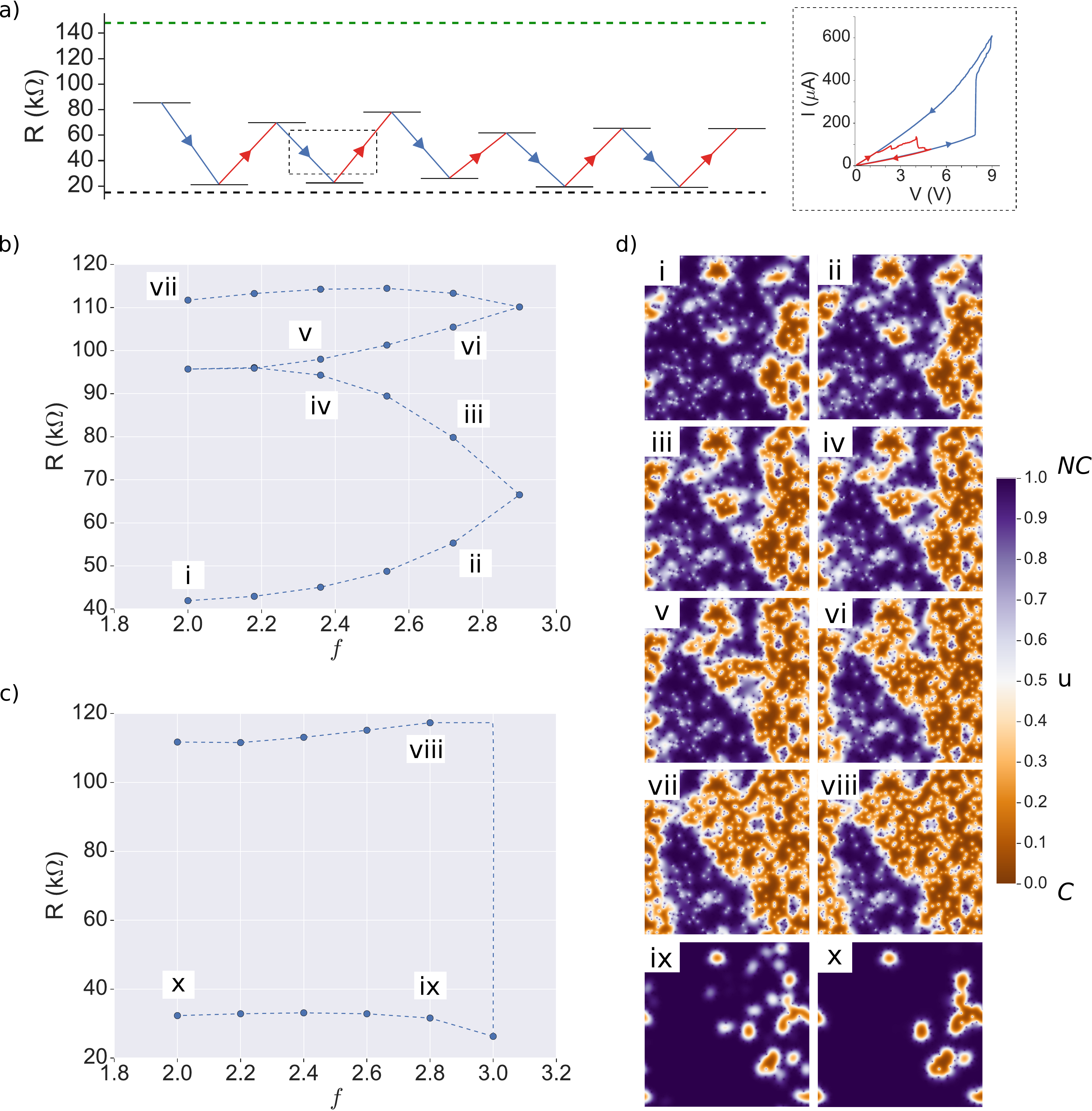}
\caption[TDGL simulations]{TDGL simulations of bidirectional resistance switching. (a) (Left) Experimental data demonstrating the bidirectional switching of the bulk resistance of an ultrathin sample. The resistance is repeatedly toggled between a low resistance state comparable to that obtained when cooling the sample from above the NC-C phase transition (black dashed line), and a high resistance intermediate state. The resistance of the state reached by warming the sample from $77 \textrm{ K}$ is given by the green dashed line. Driving to the high resistance state (red) was performed by applying an AC bias at $17.8 \textrm{ Hz}$, while driving to the low resistance state (blue) was performed by applying a DC bias. (Right) The bidirectional switching outlined by the dashed box. For the red trace, the voltage applied corresponds to $V_{rms}$. (b) A simulated electrical driving sweep using the TDGL model described in the text. Starting from a state predominantly in the NC phase, $f$ is increased, resulting in the conversion of some NC regions to C domains. In the simulated sweep, $f$ plays the role of the applied voltage in the electrical driving. The Roman numerals labelling the points in the sweep refer to the corresponding microstructures in (d). (c) Continuing the sweep from (b), starting from a microstructure that is predominantly C, $f$ is swept until $f = f_c = 3.0$, at which point C regions are converted to NC domains that form around clusters of pinning centers. (d) Examples of the simulated microstructures generated in sweeps (b) and (c). $u=0$ represents the C phase, while $u=1$ is the NC phase.}
\label{fig:drivingSims}
\end{figure*}

\begin{figure*}
\centering
\includegraphics[width=0.8\linewidth]{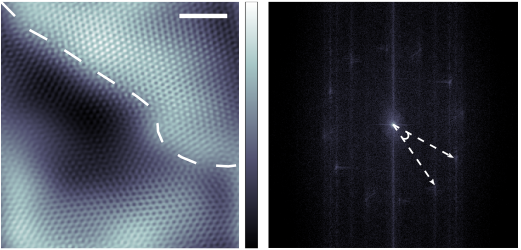}
\caption[Chiral states in 1T-\ch{TaS2}.]{(Left) A mirror twin boundary is observed in an ultrathin sample of 1T-\ch{TaS2}. $\textrm{Scale bar} = 8 \unit{nm}$, $V_t = 0.44 \textrm{ V}$, $I_t = 0.75 \textrm{ nA}$. (Right) FFT of the same region. Twelve peaks are visible indicating the coexistence of the two different chiral orientations of the CDW in 1T-\ch{TaS2}. The angle between the two vectors indicated is equal to $\sim 26.3^{\circ}$.}
\label{fig:SIchiral}
\end{figure*}

The data collected in this work does not preclude the possibility that the driving of the inhomogeneous electronic state involves a continuous evolution towards the C phase, such as through a modification of the domain period in NC-like regions with the creation or annihilation of domain walls. However, many of the features observed in the bidirectional resistance switching of ultrathin 1T-\ch{TaS2} can be explained by the system being in a phase separated state. For a flake in a phase separated state, the bulk resistivity is determined by the arrangement and relative volume fractions of the NC-like and C-like domains, with the initial volume fractions being set by the temperature of the system. Given the nonuniformity of the charge ordering in a phase separated state, an applied voltage drops asymmetrically across the sample. When the flake is in the low resistance state, there exists a percolative NC path, and the voltage drops across the NC region spanning the sample. The applied electric field stabilizes the C phase relative to the NC phase, causing the C state volume fraction to increase until the percolative NC path no longer exists and the resistance increases. With an increase in the C phase volume fraction, less of the applied voltage drops across NC domains, causing the driving to halt at an intermediate resistance value (Figure~\ref{fig:drivingSims}a). As the voltage is increased further, current flow through the C domains increases until breakdown occurs and the metallic percolative path is again formed, causing the resistance to decrease. With the breakdown of the C domains, the voltage again drops across an NC path, and consequently it is possible to observe a transition back into a more resistive state (Figure SI~\ref{fig:SIDC}c). This result was previously seen in the literature where the authors were unable to drive the sample back to the low resistance value.\cite{patel2020photocurrent} However, Figure~\ref{fig:drivingSims}a demonstrates that we are able to fully toggle between the saturation resistance level and the resistance minimum by using a DC bias when driving from the high to the low resistance state, and an AC signal when driving the sample in the opposite direction.

We aim to provide some qualitative insight into the bidirectional switching using the picture described above and modeling the system using time-dependent Ginzburg-Landau theory (TDGL). In TDGL the evolution of an order parameter $u$ is given by:\cite{hohenberg1977theory}
\begin{equation}
\frac{\partial u}{\partial t} = - \Gamma \frac{\partial F}{\partial u}
\label{eqn:TDGLeqn}
\end{equation}
where $\Gamma$ is a damping parameter. Starting from a Landau free energy $F$ with minima at $u = 0$ (the C phase) and $u = 1$ (the NC phase), we introduce the phenomenological parameter $f$ to tune the stability of the C phase relative to the NC phase by tilting the free energy potential:
\begin{equation}
F(u) = \frac{a^2}{2} u^2 (1-u)^2 + f u^2 (3 - 2u) + \frac{1}{2} ( \nabla u)^2,
\label{eqn:freeenergy}
\end{equation}
where $a$ controls the magnitude of the energy barrier between the C phase and the NC phase. Here we use a single order parameter rather than three parameters as is often done in describing a triple CDW,\cite{mcmillan1976theory,nakanishi1977nearly} and ignore discommensurations.

A phase separated structure is produced only upon introducing disorder into the system in the form of pinning centers, point defects where the order parameter is fixed to $u = 1$, the NC phase. In the quasi-2D limit, defect sites of this kind would play a significant role in determining the properties of 1T-\ch{TaS2}, compared to in bulk samples of the material.\cite{tsen2015structure} In the generated microstructure, NC domains form around clusters of defect sites (Figure~\ref{fig:drivingSims}d), and the relative volume fractions of the C and NC phases are determined based on the value of $f$ and the number of pinning centers in the simulation. By introducing a limiting value $u_{lim} = 0.5$ below which the order parameter is depinned from the defect sites, meaning it is not reset to $u = 1$, we realize bistability of the C and NC phase for a given value of $f$: the relative volume fractions of the C and NC phases become dependent upon the initial state of the system in addition to the tilt of the potential. An additional threshold parameter, $f_c$, is used to set the point at which the C domains are converted to NC. We calculate the resistance of each microstructure by discretizing the image into a 2D resistor network, and solving for an equivalent resistance.\cite{frank1988highly}

Utilizing the model described above, we qualitatively reproduce the main features observed during the electrical driving. In particular, we demonstrate how the evolution of a phase separated state can give rise to microstructures with bulk resistivities that lie between the values associated with the homogeneous CDW phases. A linear voltage sweep is simulated by a linear sweep of $f$.
Figure~\ref{fig:drivingSims}b illustrates a simulated sweep starting from a microstructure that is predominantly in the NC phase. Examples of the simulated microstructures generated during the sweep are given in Figure~\ref{fig:drivingSims}d, panels (i-vii). As $f$ is increased, some of the pinning centers are depinned, causing the volume fraction of the C phase (orange domains in Figure~\ref{fig:drivingSims}d) to increase, and resulting in an increase in the bulk resistance. When $f$ is swept back to its initial value, the state the system returns to is different than its initial state, as the steady-state solution depends on both the value of $f$ and on the history of the material. Figure~\ref{fig:drivingSims}c demonstrates a sweep starting from the predominantly C phase from Figure~\ref{fig:drivingSims}d, panel (vii). Once $f$ is increased to $f_c$, NC domains form around clusters of pinning centres (Figure~\ref{fig:drivingSims}d panel (ix)). The state of the system when $f$ is returned to its initial value has a significantly larger volume fraction of NC regions compared to the initial microstructure, and thus a lower resistance.

In summary, we used STM to image ultrathin flakes of 1T-\ch{TaS2} in the hysteresis region of the NC-C phase transition. The exfoliated flakes, with lateral dimensions $< 10 \textrm{ } \mu \textrm{m}$, were integrated into a device design that allows for the observation of changes in the electronic structure of the material that accompany changes in bulk electrical properties. Within the hysteresis region of the NC-C phase transition, we found ultrathin 1T-\ch{TaS2} to exist in a state with electronic inhomogeneity, indicative of the competition between different equilibrium states. This result highlights the complexity in the phase diagram of 1T-\ch{TaS2} and demonstrates the importance of understanding the characteristics of its phase competition morphologies in order to identify potential applications for this material.

\bibliography{references}

\begin{acknowledgments}

The authors thank Neil Curson and Joseph Salfi for helpful discussions. This research was undertaken thanks in part to funding from the Canada First Research Excellence Fund. J.O. acknowledges the support by Georg H. Endress Foundation. JJG, X.L., WJL and Y.P.S. thank the financial supports from the National Natural Science Foundation of China under Contract Nos. 11674326, 11874357, the Joint Funds of the National Natural Science Foundation of China, the Chinese Academy of Sciences' Large-Scale Scientific Facility under Contract Nos. U1832141, U1932217, and U2032215, the Key Research Program of Frontier Sciences, CAS (No. QYZDB-SSW-SLH015), the uses with Excellence and Scientific Research Grant of Hefei Science Center of CAS (No.2018HSC-UE011). A.W.T. acknowledges support from the US Army Research Office (W911NF-21-2-0136), Ontario Early Researcher Award (ER17-13-199), and the Natural Science and Engineering Research Council of Canada (NSERC) (RGPIN-2017-03815).

\end{acknowledgments}

\newpage

\section{Supplementary Material for ``Observation and manipulation of a phase separated state in a charge density wave material''}
\subsection{Fabrication Details}
\subsubsection{Synthesis of \texorpdfstring{1T-\ch{TaS2}}{1T-TaS2}}
Single crystals of 1T-\ch{TaS2} were grown by chemical vapor transport with iodine as the transport agent. High-purity \ch{Ta} ($3.5$ N) and \ch{S} ($3.5$ N) were mixed in chemical stoichiometry and heated at $850^{\circ} \textrm{C}$ for $4$ days in an evacuated quartz tube. The harvested \ch{TaS2} powders and iodine (density: $5 \textrm{ mg/cm}^3$) were then sealed in another quartz tube and heated for $2$ weeks in a two-zone furnace, where the source and growth zones were held at $900^{\circ} \textrm{C}$ and $800^{\circ} \textrm{C}$, respectively. The tubes were then rapidly quenched in cold water to retain the 1T phase.

\subsubsection{Electrode fabrication}
Electrodes were fabricated on an \ch{SiO2} ($285$ nm)/\ch{Si} substrate. Standard electron beam lithography and e-beam metal deposition techniques were used to fabricate $20 \textrm{ }\mu \textrm{m} \times 20 \textrm{ }\mu \textrm{m}$ pads of \ch{Au} ($12$ nm)/\ch{Ti} ($3$ nm). An array of $1 \textrm{ }\mu \textrm{m}$ size markers (\ch{Au} ($7$ nm)/\ch{Ti} ($3$ nm)) spaced $2 \textrm{ }\mu \textrm{m}$ apart was deposited on top of the pads. The pads were contacted by larger $120 \textrm{ }\mu \textrm{m} \times 55 \textrm{ }\mu \textrm{m}$ contacts of \ch{Au} ($60$ nm)/\ch{Ti} ($10$ nm), fabricated using mask-less photolithography and e-beam deposition.

\subsubsection{2D material transfer procedure}
To prevent oxidation of the 1T-\ch{TaS2} flakes the exfoliation and transfer of 2D materials was done inside a \ch{N2} filled glovebox with \ch{O2} and \ch{H2O} partial pressures below $0.1$ ppm. The materials were exfoliated and transferred sequentially on top of each other using the following method:
\begin{enumerate}
\item A bulk crystal of the 2D material is exfoliated onto a piece of Scotch tape.
\item A polypropylene carbonate (PPC) thin film is prepared by spin coating ($1600$ RPM, $30$ sec) PPC in an anisole solution ($15 \%$ PPC by weight) on a glass cover slide. The Scotch tape with the exfoliated flakes is pressed onto the thin film of PPC and then peeled off.
\item The PPC thin film is transferred onto a polydimethylsiloxane (PDMS) stamp sitting on a glass slide.
\item A home-built setup utilizing a motorized arm attached to a microscope is used to identify a suitable flake on the PPC film.
\item The selected flake is aligned with the fabricated electrodes as desired and brought into contact with the substrate. The substrate is then heated to $90^{\circ} \textrm{C}$, causing the PPC to melt and peal off of the glass slide at the point of contact, transferring the flake and PPC film to the substrate.
\item The heater is switched off and the electrodes with the transferred flake and PPC are put into a vacuum chamber for $30$ min. before the PPC is washed away in a chloroform bath, leaving only the flake on the substrate.
\item The sample is then rinsed with acetone and isopropanol.
\end{enumerate}

\subsubsection{STM measurements}
All STM measurements were performed with an Omicron LT-STM outfitted with four electrical contacts. STM topography was taken in constant-current mode with the bias applied to the tip. The samples were heated to $115^{\circ} \textrm{C}$ prior to being transferred to the STM chamber.

Ultrathin flakes of 1T-\ch{TaS2} are difficult to image with STM due to the instability of the material in air as well as the requirement that the sample not be heated above $500 \unit{K}$, as that would cause the TMD to undergo a structural phase transition to the 2H polytype. However, despite the processing difficulties described above and the exposure of the samples to various chemicals during fabrication, the exfoliated flakes we measured remained sufficiently free of contaminants to resolve the CDW superstructures characteristic of the bulk phases of 1T-\ch{TaS2}.

\subsection{Chiral states}
As 1T-\ch{TaS2} undergoes a phase transition from the incommensurate (IC) charge density wave (CDW) phase to the nearly commensurate (NC) CDW phase two chiral orientations are possible. In the IC phase the CDW wave vector is oriented along the atomic lattice wave vector while in the NC phase the CDW is rotated away from the atomic lattice at an angle that is dependent on the temperature of the system ($\sim 12^{\circ} - 13.9^{\circ}$). The two possible chiral orientations correspond to whether the CDW is rotated either clockwise or counterclockwise.

Both orientations are equivalent and have been observed to exist within the same sample.\cite{zong2018ultrafast} Furthermore, interconversion between the different orientations has been shown via femtosecond laser pulses.\cite{zong2018ultrafast}. We observe the coexistence of the two chiral phases separated by a mirror twin boundary.

\subsection{Further electrical characterization}
In this section, transport measurements are presented for two 1T-\ch{TaS2} samples of a similar thickness, $\sim 8 \textrm{ nm}$, hereby called Sample A, and Sample B. Sample A was the device imaged in Figures 2 and 3 in the main text, while Sample B is the device driven in Figure 4d and 4e.

\begin{figure}
\centering
\includegraphics[width=0.95\linewidth]{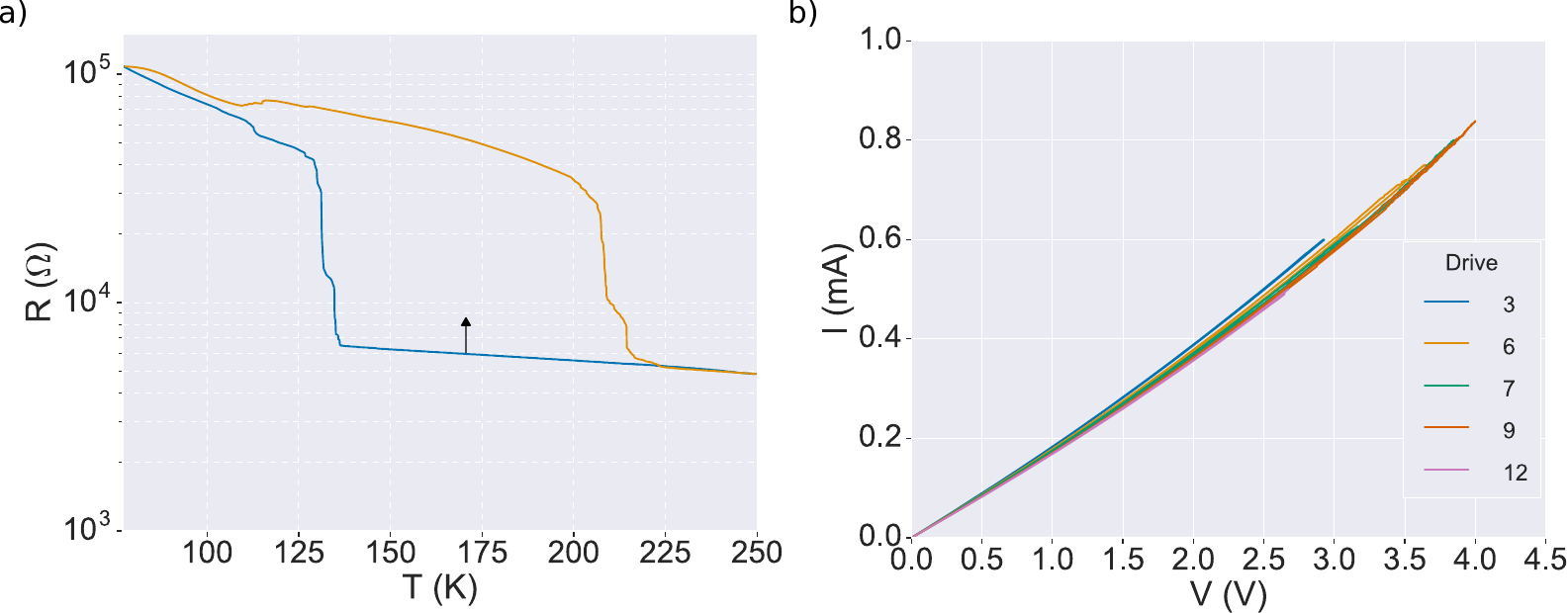}
\caption[Electrical driving of 1T-\ch{TaS2}.]{(a) Temperature-dependent resistance of Sample A. The traces were taken after the sample was driven by applying a lateral bias across the flake. The orange trace was measured upon warming the sample from $77 \textrm{ K}$ while the blue trace was collected upon cooling the sample. (b) The IV traces measured while driving the sample as depicted in Figure 3 of the main text. In this instance the resistance only increased by $\sim 1\%$.}
\label{fig:SITASTM5}
\end{figure}

The resistance of the device was measured using a four-terminal, constant current circuit that does not exclude the contact resistance. For all DC voltage sweeps, a Keithley Source Measurement Unit (Keithley 2401 or Keithley 2450) was used to apply the bias voltage and to measure the resulting current. For the traces plotted in Figure~\ref{fig:SITASTM5}, the DC sweep rate was $0.05 \textrm{ V/s}$. AC measurements were performed with a SR830 lockin, with a signal frequency of $17.8 \textrm{ Hz}$. The circuit included a current divider in order to protect the lockin from excessive current.

\begin{figure*}
\centering
\includegraphics[width=0.65\linewidth]{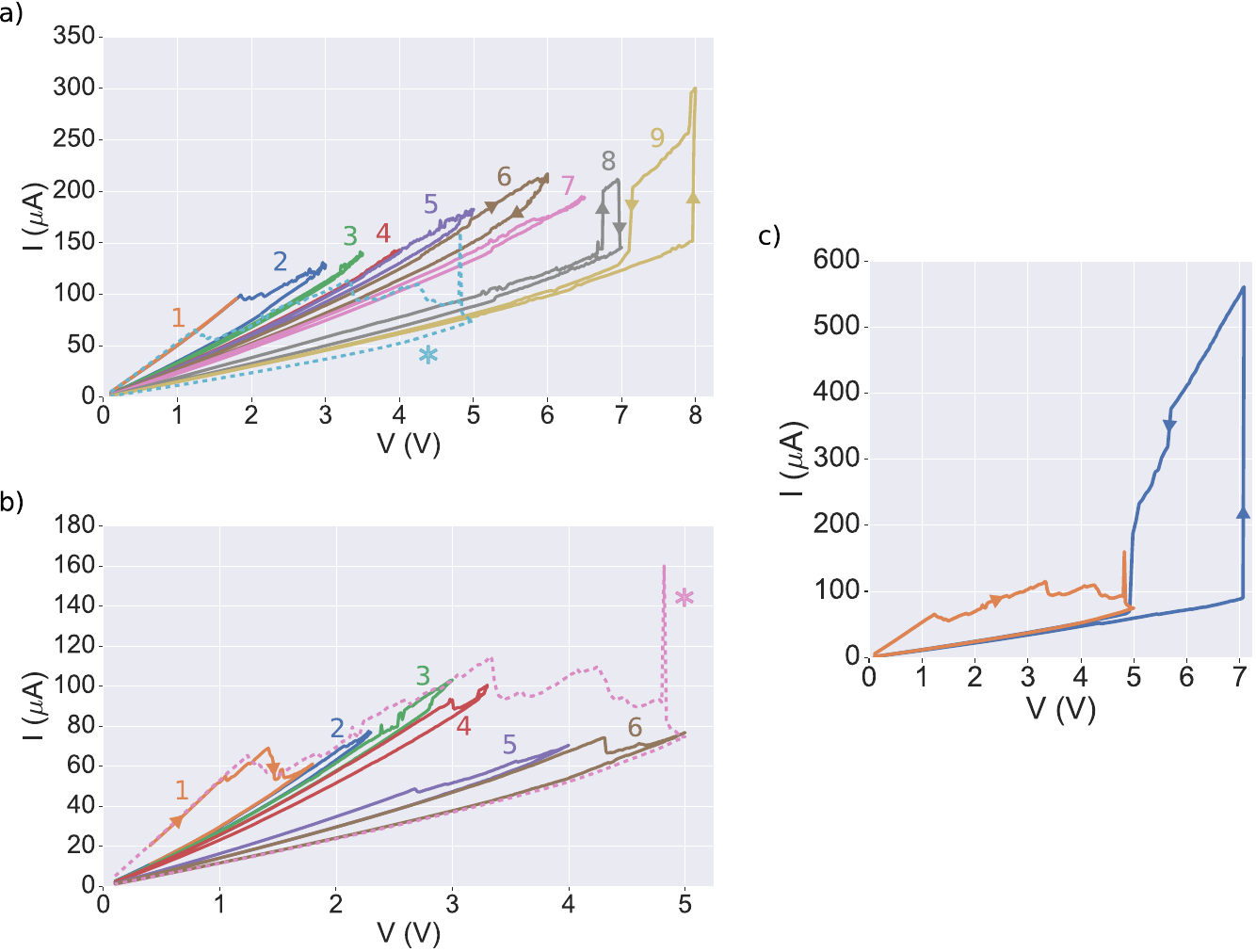}
\caption[Comparison between driving Sample B with a DC bias and with an AC bias.]{(a) Sample B driven by applying a lateral DC bias. The sample sat overnight between drive 7 and drive 8. The trace labelled with an asterisk is a drive performed by applying a lateral AC bias (see Figure~\ref{fig:SIACDC}b panel i) and indicates the high and low resistance states of the sample. (b) Sample B driven by applying a lateral AC bias. The sample sat overnight between drive 4 and drive 5. The trace labelled with an asterisk is the same as the AC drive in (a) illustrating how the high and low resistance states do not vary whether a DC or AC bias is used. (c) Attempting to drive Sample B further by going to a greater AC voltage. Using only a lateral AC bias, we were unable to drive the sample from the high resistance state to the low resistance state. As the voltage was swept back to the $0 \textrm{ V}$ the sample returned to the high resistance state. In all plots, the direction of the voltage sweep coincides with a transition from a low to a high resistance state (e.g. as illustrated in Drive $1$ of (b)), unless otherwise indicated by arrows.}
\label{fig:SIDC}
\end{figure*}

As evidenced in Figure~\ref{fig:drivingSims}b from the main text, multiple sweeps to the same voltage value will continuously manipulate the state of the material. Assuming the maximum voltage set is less than the voltage at which breakdown of the C domains occurs, the material will be driven to some intermediate state with a higher resistance compared to the initial state. With this consideration in mind, applying a lateral AC bias across the sample will drive the material more efficiently than a DC bias, as when using an AC signal multiple sweeps are performed. Figure~\ref{fig:SIDC} demonstrates this behavior.
\begin{figure*}
\centering
\includegraphics[width=0.75\linewidth]{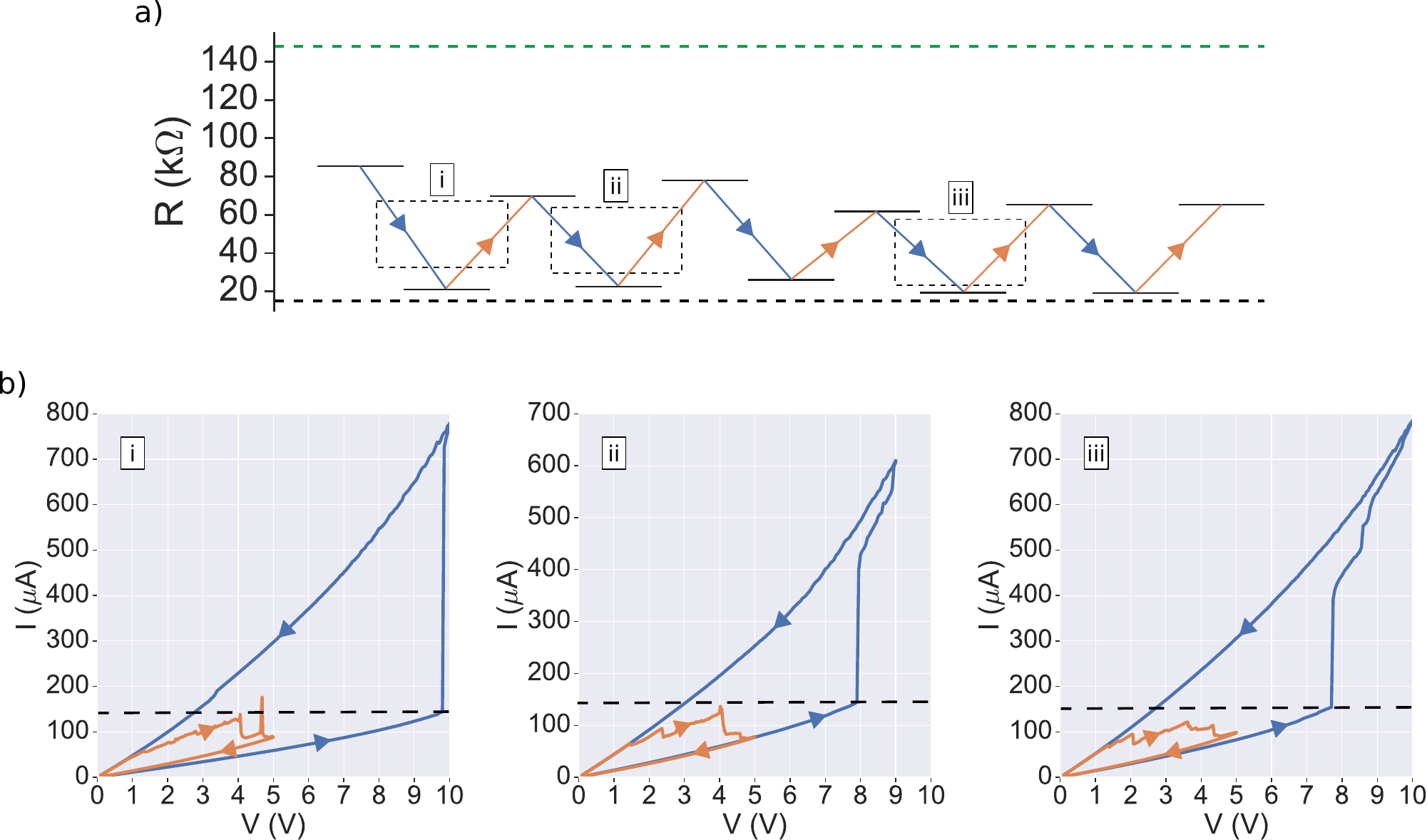}
\caption[Bidirectional switching of 1T-\ch{TaS2}.]{(a) Experimental data demonstrating the bidirectional switching of the bulk resistance of the sample. The resistance is repeatedly toggled between a low resistance state comparable to that obtained when cooling the sample from above the NC-C phase transition (black dashed line), and a high resistance intermediate state. The resistance of the state reached by warming the sample from $77 \textrm{ K}$ is given by the green dashed line. (b) The individual traces outlined in the dashed boxes in (a). Driving to the high resistance states (orange curves) was performed by applying an AC bias while driving to the low resistance state (blue curves) was performed by applying a DC bias. The critical current at which the transition from the high to the low resistance state occurs is approximately equal in all three plots, as indicated by the dashed black line.}
\label{fig:SIACDC}
\end{figure*}
Using an AC bias at $17.8 \textrm{ Hz}$, the sample is driven to the saturation resistance level more directly, and at lower voltages compared to when a DC bias is used. Whether there exists an alternative explanation for the enhancement observed when using an AC voltage, one that is based upon the underlying microscopic mechanism that facilitates the electrical driving, is beyond the scope of this work.

\subsection{Time-dependent Ginzburg-Landau theory}
We aim to provide some qualitative insight into the phase separation dynamics using time-dependent Ginzburg-Landau (TDGL) theory. While the main results of the simulations were provided in the main text, this section contains further details of the modeling.

The governing equation in TDGL is:\cite{hohenberg1977theory}
\begin{equation}
\frac{\partial u}{\partial t} = - \Gamma \frac{\partial F}{\partial u}
\label{SI-eqn:TDGL}
\end{equation}
where $\Gamma$ is a damping parameter. Starting from a Landau free energy $F$ with minima at $u = 0$ (the C phase) and $u = 1$ (the NC phase) we introduce the phenomenological parameter $f$ to tune the stability of the C phase relative to the NC phase by tilting the free energy potential:
\begin{equation}
F(u) = \frac{a^2}{2} u^2 (1-u)^2 + f u^2 (3 - 2u) + \frac{1}{2} ( \nabla u)^2
\end{equation}
where $a$ controls the magnitude of the energy barrier between the C phase and the NC phase, and the energy is measured in units of the elastic energy.

\begin{figure*}
\centering
\includegraphics[width=0.45\linewidth]{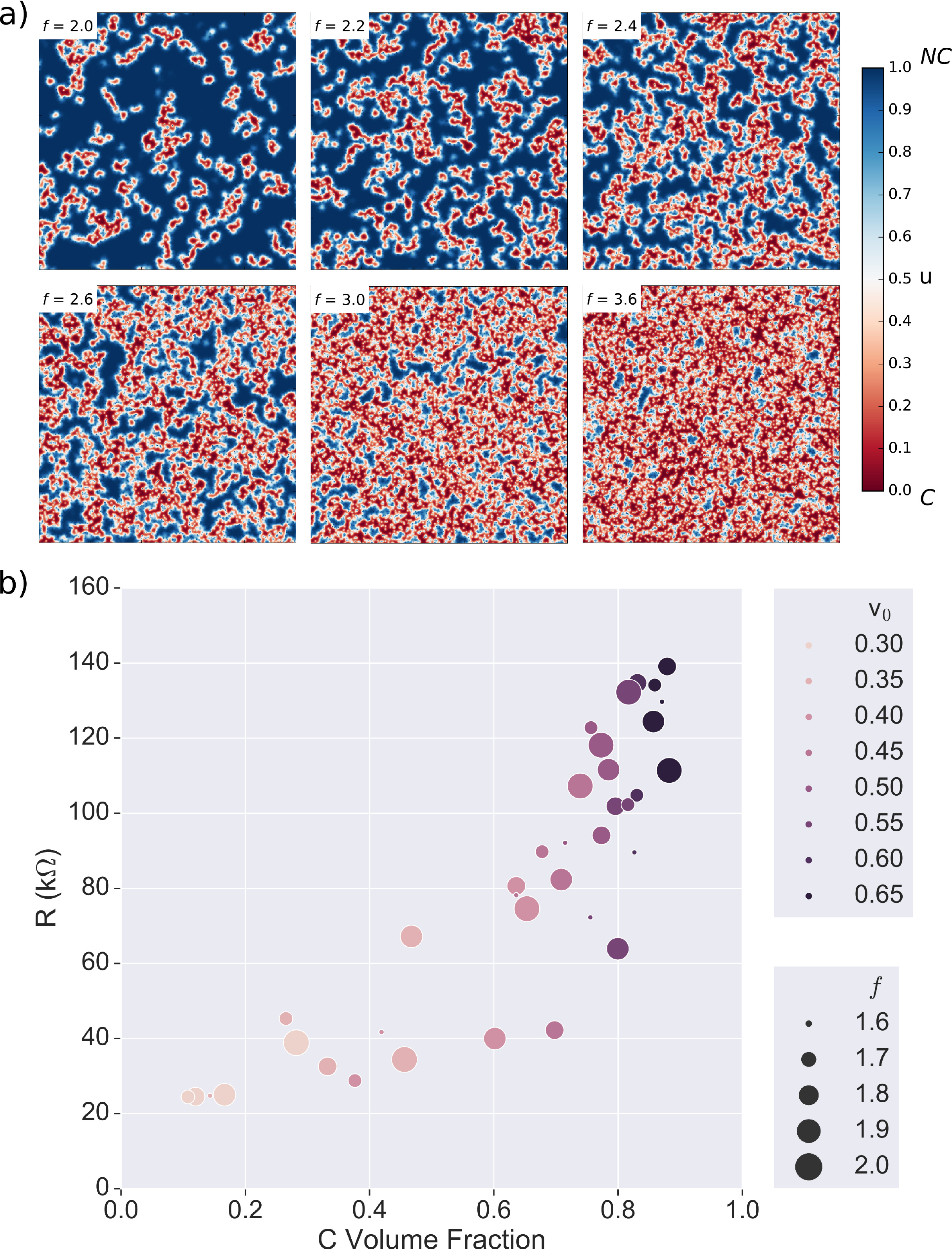}
\caption[TDGL general results.]{(a) Microstructures formed at various values of the tilting parameter $f$. (b) The resistances calculated for microstructures generated from different values of $f$ and initial C phase volume fractions ($v_0$).}
\label{fig:SImicrostructures}
\end{figure*}

Starting from a random, inhomogeneous microstructure, the system is allowed to evolve according to Equation~\ref{SI-eqn:TDGL}. Without introducing disorder, the system will form either the homogeneous C or NC phase, depending on the value of the tilt parameter, $f$. However, if disorder is introduced into the model in the form of pinning centers, impurity sites where the order parameter is reset to $u=1$ at the end of each time step, a phase separated structure containing intertwined C and NC phase domains is generated. The NC domains are situated around clusters of pinning centers, and the relative volume fractions of the two phases depend on both $f$ and the total number of impurity sites introduced (Figure~\ref{fig:SImicrostructures}a). A 5-point Laplacian discretization is used in solving Equation~\ref{SI-eqn:TDGL}. For all of the presented simulations, the impurity concentration was set to $0.01$ and the system was allowed to equilibrate over $10000$ time steps at each value of $f$.

\begin{figure*}
\centering
\includegraphics[width=0.85\linewidth]{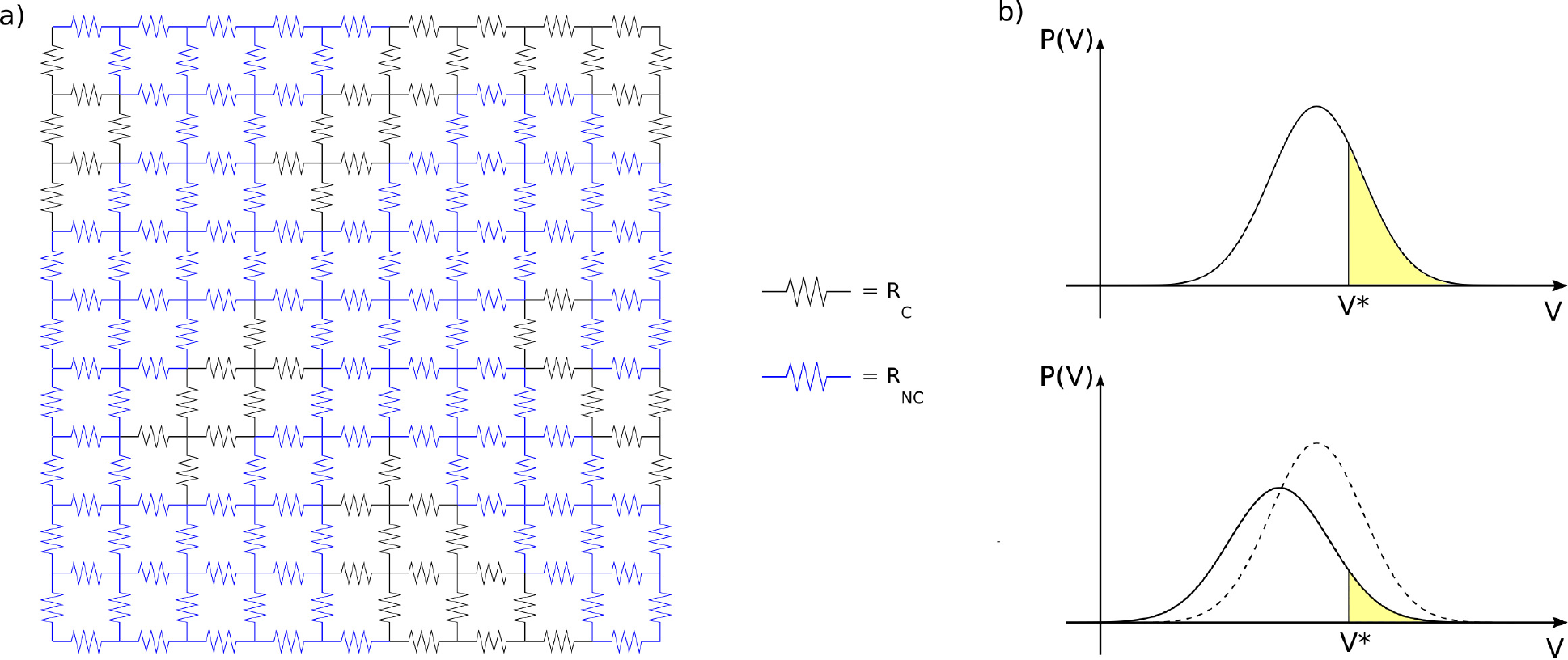}
\caption[2D resistor network.]{(a) The system is discretized and modeled as a 2D resistor network. The resistance of the microstructure is calculated using the standard $Y-\nabla$ transformation.\cite{frank1988highly} (b) The voltage dropping across the NC nodes will follow some distribution $P(V)$ (left). NC nodes with a value beyond the threshold value $V*$ are converted to C, changing the distribution of the voltage drops (right).}
\label{fig:SImodelingSchematic}
\end{figure*}

In order to simulate the electrical driving we model the system as a 2D resistor network with NC and C nodes. Initially, the system is considered to be in a state with a small C phase volume fraction, as depicted in Figure~\ref{fig:SImodelingSchematic}a.
Due to the coexistence of the C and NC phase domains, the voltage applied across the sample will not drop homogeneously but will depend on the geometric arrangement of the microstructure. Consequently, the voltage drop across the complete set of NC regions will have some distribution $P(V)$. An applied electric field stabilizes the C phase relative to the NC phase. With respect to the model, this effect corresponds to an increase in the parameter $f$. As a voltage is applied to the sample, the voltage drop across some of the NC nodes exceeds a certain critical value $V*$ resulting in those NC nodes in the resistor network to become C. When this conversion occurs the distribution of the voltage across the NC nodes changes (Figure~\ref{fig:SImodelingSchematic}b).

The general picture described above is approximated in the simulations by introducing a limiting value $u_{lim} = 0.5$ below which the order parameter is depinned from an impurity site, meaning it is not reset to $u = 1$ at the end of each time step. With this modification, the final microstructure generated from the time evolution of the system becomes dependent on the initial C phase volume fraction, $v_0$, in addition to the tilting parameter and the number of pinning centers. For small values of $v_0$ ($v_0 \leq 0.40)$, the system evolves to a percolated NC phase, with a resistivity comparable to that of the homogeneous NC phase, while for larger $v_0$, the system goes to the C phase. This result is demonstrated in Figure~\ref{fig:SImicrostructures}b, where the resistance of the generated microstructure is calculated for different sets of values of $v_0$ and $f$. The driving model is completed by adding a second set of threshold parameters, $f_c$ and $u_c$, to control the conversion of C nodes to NC nodes. When $f = f_c$, nodes where the order parameter is greater than $u_c$ are converted to $u = 1$, the NC phase.

The model described above does not explicitly allow for the rearrangement of the NC and C domains as the location of the pinning centres essentially fixes the geometric arrangement of the microstructure. Thermal noise and Joule heating effects stemming from the flow of charge carriers are not included, but would result in a more complete picture that might allow for the possibility of the rearrangement of the microstructure. Despite this limitation, the model does recreate many of the essential features of the observed driving.

\end{document}